\RequirePackage{lineno}
\documentclass[aps,showpacs,twocolumn,nofootinbib]{revtex4}
\usepackage{graphicx}
\usepackage{adjustbox}
\usepackage{dcolumn}
\usepackage{bm}
\usepackage{multirow}
\usepackage{amsmath}
\usepackage[letterpaper, hmargin = {2cm}, vmargin = {2cm}]{geometry}

\usepackage[final]{changes}

\begin{document}


\preprint{PRC draft of 33Mg KO}

\title{ Analysis of  Spectroscopic Factors in $^{11}$Be and $^{12}$Be\\
 in the Nilsson Strong Coupling Limit }

\author{A.~O.~Macchiavelli, H.~L.~Crawford, C.~M.~Campbell, R.~M.~Clark, M.~Cromaz, P.~Fallon, M.~D.~Jones, I.~Y.~Lee,  and M.~Salathe}
\affiliation{Nuclear Science Division, Lawrence Berkeley National Laboratory, Berkeley, CA 94720, USA}
\date{\today}
 
\begin{abstract}
Spectroscopic factors  in $^{10}$Be, $^{11}$Be  and $^{12}$Be, extracted from $(d,p)$, one neutron knockout, and $(p,d)$ reactions are interpreted within the rotational model.  
Assuming that the ground state and first excited state of $^{11}$Be can be associated with the  $\frac{1}{2}[220]$ and  $\frac{1}{2}[101]$ Nilsson levels, the strong coupling limit gives simple expressions that relate the amplitudes of these wavefunctions (in the spherical basis) with the measured cross-sections and derived spectroscopic factors.  
We obtain good agreement with both the measured magnetic moment of the ground state in $^{11}$Be and the reaction data.
\end{abstract}


\maketitle

\section{Introduction}

The lightest example of a so-called  ``Island of Inversion'' is that at $N$ = 8, where the removal of $p_{3/2}$ protons from $^{14}$C results in a quenching of the 
$N$=8 shell gap~\cite{Talmi, Sor08, Heyde1, Heyde2}.   This is evident with the sudden drop of the $E(2^+)$ energy in $^{12}$Be relative to the neighboring even-even isotopes, and the inversion of the ground state of 
$^{11}$Be from the expected 1/2$^-$   to the observed positive parity  1/2$^{+}$  state.

The underlying physics of such inversions is now rather well understood,  with the neutron-proton interaction playing an important role~\cite{Sor08, Heyde1, Heyde2}.
Changes in the monopole average of the central and spin-isospin components of this interaction when
protons are removed induces changes in the neutron effective single-particle energies (ESPEs) with the effect of eroding the expected shell closure~\cite{Ots01}.
Taking advantage of this erosion, the quadrupole interaction takes over and drives the system to deform.
 Specifically for the case at $N$=8, the inversion is driven by the combined effects of the $V^{\pi\nu}_{p_{3/2}, p_{1/2}}$ interaction, and the lowering of the $s_{1/2}$ orbit due to weak binding~\cite{Calem}.  

Given the key role of deformation, it is  of interest to understand the structure of neutron-rich Be isotopes within the Nilsson model~\cite{Sven, Rag}.
In fact,  Bohr and Mottelson~\cite{BM} argued for the role of deformation to explain the inversion of the 
$1/2^{+}$ and the $1/2^{-}$ states in $^{11}$Be. Building on that premise,  Hamamoto and Shimoura~\cite{Ikuko} presented a detailed interpretation of 
energy levels and available electromagnetic data on $^{11}$Be and $^{12}$Be  in terms of single-particle
motion in a deformed potential, using weakly bound one-particle wavefunctions  calculated 
with a deformed Woods-Saxon (WS) potential instead of the standard  harmonic-oscillator potential~\cite{Larsson}.
The role of deformation in the Be isotopes and its relation to clustering phenomena has also been extensively discussed~\cite{VonO}.

In this work we analyze spectroscopic factors,  obtained from studies of the $^{11}$Be$(d,p)$$^{12}$Be~\cite{Ritu,Johansen},  $^{10}$Be$(d,p)$$^{11}$Be~\cite{Auton, dpbe10}, $^{12}$Be one neutron knockout ($-1n$)~\cite{Navin,Pain} and
$^{11}$Be$(p,d)$$^{10}$Be~\cite{Winfield} reactions, in the Nilsson strong coupling limit. We use the formalism reviewed in Ref.~\cite{Elbek}, which we have recently applied to the $N$=20 Island of Inversion~\cite{aom1}.

\section{The Method}

Following  Refs.~\cite{BM, Ikuko} we associate the $1/2^{+}$ and the $1/2^{-}$ states in $^{11}$Be to the Nilsson levels $\frac{1}{2}[220]$ and  $\frac{1}{2}[101]$ respectively.
In the spherical $|j,\ell\rangle$ basis these wavefunctions take the form:

\begin{equation}
|\tfrac{1}{2}[220]\rangle=  C_{1/2,0} |s_{1/2}\rangle+ C_{3/2,2} |d_{3/2}\rangle+  C_{5/2,2}|d_{5/2}\rangle
\label{eq:eq1}
\end{equation}

\begin{equation}
|\tfrac{1}{2}[101]\rangle=  C_{1/2,1} |p_{1/2}\rangle+ C_{3/2,1} |p_{3/2}\rangle 
\label{eq:eq2}
\end{equation}
where $C_{j,l}$ are the associated Nilsson wavefunction amplitudes.

For transfer reactions, such as $(d,p)$,  the spectroscopic factors ($S_{i,f}$) from an initial ground state $|I_i K_i\rangle$ to a final state $| I_f K_f \rangle$ can be written in terms of the Nilsson amplitudes:
\begin{equation}
S_{i, f  }= \frac{ (2I_i+1)}{(2I_f+1)} g^{2}\langle I_{i}jK_{i}\Delta K | I_{f}K_{f}\rangle^{2}C_{j,\ell}^{2} \langle\phi_f|\phi_i\rangle^2 
\label{eq:eq2}
\end{equation}
where $g^{2}$ = 2 if $I_{i}$ = 0 or $K_{f}$ = 0 and $g^{2}$ = 1 otherwise,
and $\langle\phi_f|\phi_i\rangle$ represents the core overlap between the initial and final states.  A similar expression, without the spin factors, applies to the cases of 1n-KO and $(p,d)$.

Finally,  we consider the final 0$^{+}$ states in  $^{12}$Be as superpositions of the neutron states in Eqs. (1,2)~\cite{Ikuko}: 
\begin{equation}
|0^+_1\rangle=  \alpha | \nu_1 \bar\nu_1\rangle + \beta | \nu_2 \bar\nu_2\rangle  \nonumber
\end{equation}
\begin{equation}
|0^+_2\rangle=  -\beta| \nu_1 \bar\nu_1\rangle + \alpha| \nu_2 \bar\nu_2\rangle 
\label{eq:eq3}
\end{equation}
\noindent
where $ | \nu_1\rangle$  and  $ | \nu_2\rangle$  are the neutron states in Eq.~(1) and Eq.~(2) respectively and $\bar\nu$ indicates the time-reverse orbit.
The  $|2^+_1\rangle$ is associated with the 2$^+$ member of the rotational band built on the $|0^+_1\rangle$ state. 

\section{Results}

With the established framework for our calculations, we can derive specific formulae relating the Nilsson amplitudes $C_{j,l}$ to the experimental spectroscopic factors for the reactions considered here. The relations follow directly from Eqs. (1-4) and are given below for the four specific cases.  
 
\subsection{ $^{11}$Be$(d,p)$$^{12}$Be}

\noindent
For this first case we start from the  $^{11}$Be 1/2$^{+}$ ground state, and consider transfer of a single neutron in $(d,p)$ to populate the 0$^{+}_{1}$, 0$^{+}_{2}$, and 2$^{+}_{1}$ states.  Following directly from Eqs.(1-4)  the relevant spectroscopic factors are
\footnote[4]{We note that in Ref.~\cite{Fortune},  the authors proposed a shell-model inspired solution to explain the spectroscopic factors data. In their analysis, 
they use simple mixed wavefunctions, naturally captured in  the Nilsson model.}:
\begin{eqnarray*}
S_{1/2^+,0^+_1} = 2 C_{1/2,0}^{2}\alpha^2
\end{eqnarray*}
\begin{eqnarray*}
S_{1/2^+,0^+_2} = 2 C_{1/2,0}^{2}\beta^2
\end{eqnarray*}
and
\begin{eqnarray*}
S_{1/2^+,2^+_1} = \frac{2}{5}(C_{3/2,2}^{2}+ C_{5/2,2}^{2})\alpha^2.
\end{eqnarray*}

\noindent

\subsection{ $^{10}$Be$(d,p)$$^{11}$Be}

\noindent
In this case, since we start from the $^{10}$Be 0$^+$ ground state, the angular momentum selection rules imposed by the Clebsch-Gordan coefficients in Eq.~(\ref{eq:eq2}) the spectroscopic factors directly project out the amplitudes of the wavefunctions
in the spectroscopic factors:
\begin{eqnarray*}
S_{0^+,1/2^+} = C_{1/2,0}^{2}
\end{eqnarray*}
and
\begin{eqnarray*}
S_{0^+,1/2^-} =  C_{1/2,1}^{2}.
\end{eqnarray*}
It is worth noting that this case has been studied in the particle-vibration coupling and deformed-core-plus-neutron cluster models in Refs.~\cite{Mau, Filomena}.

\subsection{ $^{12}$Be$(-1n)$$^{11}$Be}

\noindent
The case of neutron knockout is essentially equivalent to the previous examples, but we now have the addition of the core overlaps in Eq. (4) as we consider the $K=1/2^{+}$ and $K=1/2^{-}$ final states, with spectroscopic factors given by:

\begin{eqnarray*}
S_{0_1^+,1/2^+} = 2 C_{1/2,0}^{2}\alpha^2 ;~~~~ S_{0_1^+,5/2^+} = 2 C_{5/2,2}^{2}\alpha^2
\end{eqnarray*}
and
\begin{eqnarray*}
S_{0^+_1,1/2^-}=    2 C_{1/2,1}^{2}\beta^2 ; ~~~~~   S_{0^+_1,3/2^-}= 2 C_{3/2,1}^{2}\beta^2
\end{eqnarray*}

 \subsection{$^{11}$Be$(p,d)$$^{10}$Be}

\noindent
Finally, the spectroscopic factors for the (p,d) reaction populating states in $^{10}$Be reduce to:
\begin{eqnarray*}
S_{1/2^+, 0_1^+} = C_{1/2,0}^{2}
\end{eqnarray*}
and
\begin{eqnarray*}
S_{1/2^+, 2^+_1}=  C_{5/2,2}^{2}.
\end{eqnarray*}

\begin{figure*}[ht!]
\centering
\includegraphics[width=15cm]{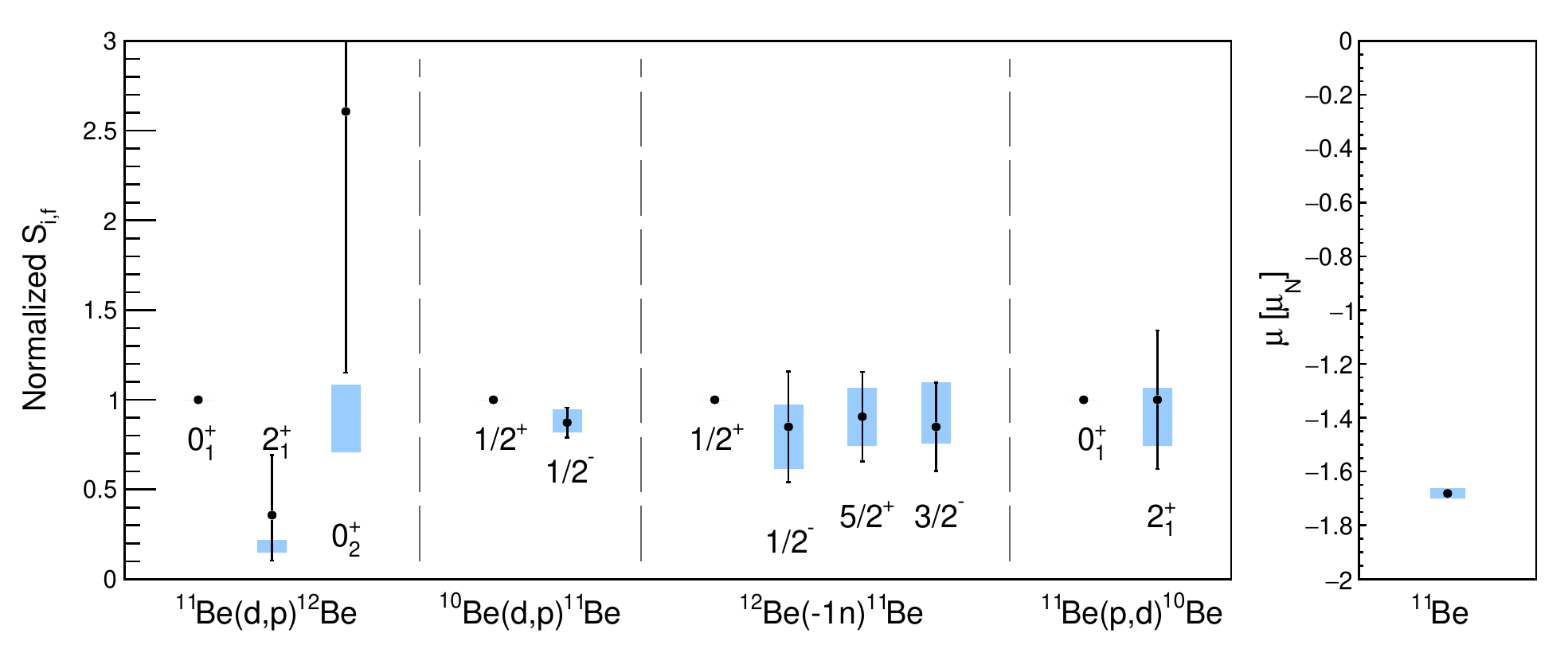}
\caption{Relative experimental spectroscopic factors and magnetic moment (data points) compared to the strong coupling limit results obtained in our analysis (blue boxes), which encompass the 1$\sigma$ confidence level in our fit. }
\label{fig:schematic}
\end{figure*}

\begin{table*}[ht]
\caption{Summary of experimental relative spectroscopic factors  in $^{10,11,12}$Be compared to the Nilsson calculations using amplitudes empirically adjusted from a weighted fit to the data.}
{\renewcommand{\arraystretch}{1.4}
\begin{tabular}{c|c|c|c|cccc|c}
\hline\hline
Initial & Final & Energy & \multirow{2}{*}{~$\ell$~}  &  \multicolumn{4}{c|}{Experimental $S_{i,f}$}  &  {Calculated $S_{i,f}$}   \\
State & State & [MeV]  &            & \cite{Ritu} &~\cite{dpbe10}  & \cite{Navin,Pain} &\cite{Winfield}&Nilsson \\
\hline 

$^{11}$Be & $^{12}$Be& & & & & &\\
$\frac{1}{2}^{+}$& 0$^{+}_1$& 0.00&0&  1 & &   & & 1\\
                          & 2$^{+}_1$& 2.11&2&    0.36$\pm0.29$  & &   & & 0.18$\pm$0.03 \\           
                          & 0$^{+}_2$& 2.24&0&   2.61$\pm1.34$ & &  & & 0.90$\pm$0.19\\         
\hline                     

$^{10}$Be & $^{11}$Be& & & & & & &\\
0$^{+}$& $\frac{1}{2}^{+}$ & 0.00&0& &1&     &&1\\
            & $\frac{1}{2}^{-}$  & 0.32&1& &0.87$\pm 0.08 $&   & & 0.88$\pm$0.06\\
 \hline         

 $^{12}$Be & $^{11}$Be& & & & & & \\
0$^{+}$& $\frac{1}{2}^{+}$ & 0.00&0&  &     & 1&&1\\
            & $\frac{1}{2}^{-}$ & 0.32&1&   &     & 0.85 $\pm 0.31$&&0.79$\pm$0.18\\
            & $\frac{5}{2}^{+}$ & 1.78&2&  &     & 0.90$\pm 0.25$&&0.91$\pm$0.16\\            
            & $\frac{3}{2}^{-}$ & 2.69&1&   &     & 0.85 $\pm 0.25$&&0.93$\pm$0.17\\
  \hline               
$^{11}$Be & $^{10}$Be& & & & & &\\
$\frac{1}{2}^{+}$& 0$^{+}_1$& 0.00&0&   & &   & 1&1\\
                          & 2$^{+}_1$& 3.4&2&    &&  & 1 $\pm0.38$&0.91$\pm$0.16\\           
                          
                           \hline\hline
\end{tabular}
}

\label{Table1}
\end{table*}

\noindent
In comparing with the experimental data (summarized in Table~\ref{Table1}) we have used the expressions above together with the condition of wavefunction normalization to empirically adjust the amplitudes of the Nilsson states in Eqs.~(1) and (2).  In addition we consider the measured magnetic moment (see Appendix) of the ground state in $^{11}$Be, $ \mu=-1.6813(5)\mu_{N}$~\cite{nndc}, as a constraint.
There are in total 12 relations connecting the experimental data to four unknown amplitudes  which we determine from a $\chi^2$-minimization procedure.  Given the possible systematic uncertainties in the determination of absolute spectroscopic factors, particularly from different experimental conditions and analysis,  we have done a weighted fit of the relative spectroscopic factor values with respect to the ground state transition for each of the data sets, and to the absolute value of the $^{11}$Be ground-state magnetic moment. 

The following wavefunctions\footnote[3]{Adopted signs follow the phases of a standard Nilsson calculation.}:
 
\begin{equation}
|\tfrac{1}{2}[220]\rangle \approx  -0.72 |s_{1/2}\rangle- 0.09|d_{3/2}\rangle+  0.69|d_{5/2}\rangle \nonumber
\label{eq:eq1}
\end{equation}
\begin{equation}
|\tfrac{1}{2}[101]\rangle \approx 0.68 |p_{1/2}\rangle+ 0.73 |p_{3/2}\rangle  \nonumber
\label{eq:eq1}
\end{equation}
\noindent
and $\alpha=0.73$ and $\beta=0.69$ are obtained.  The resulting spectroscopic factors are summarized in Table~\ref{Table1} and, with the magnetic moment, in Fig.~\ref{fig:schematic},
showing good agreement with the experimental data. The wavefunctions as well as $\alpha$ and $\beta$ are fairly consistent with those used in Ref.~\cite{Ikuko},  $\alpha$=$\beta$=0.707.  
here is continuing interest in this region of the nuclear chart, and with the availability of radioactive beams of $^{12}$Be and $^{13}$B as well as new instrumentation, further experimental work
will be carried out. With this in mind, we take the Nilsson approach a little further, and predict estimates for spectroscopic factors for the reactions $^{12}$Be$(d,p)$$^{13}$Be and $^{13}$B$(d,^3He)$$^{12}$Be which are likely to be studied in the near future.
There is some discrepancy in the literature about the low-lying level assignments of $^{13}$Be~\cite{nndc,nndc2}, but in any scenario  the $\frac{1}{2}[220]$ and  $\frac{1}{2}[101]$ Nilsson levels play a center role (as in $^{11}$Be).   The calculations are straightforward and the results are summarized in Table II.

It is also of interest to consider proton spectroscopic factors within the Nilsson scheme for $Z=5$, where the proton is expected to 
occupy the $\frac{3}{2}[101]$ level, an assignment supported by the ground state spin $3/2^-$ and measured magnetic moment in $^{13}$B, $\mu= 3.1778(5)\mu_{N}$~\cite{nndc}\footnote[5]{We calculate $\mu \approx 3.2\mu_{N}$.}.
Since the level parentage is attributed only to the $p_{3/2}$ orbit,  the spectroscopic factors depend only on the Clebsch-Gordan coefficients, 
 our predictions for the reaction $^{13}$B$(d,^3He)$$^{12}$Be  are included in Table~\ref{Table2}.

\begin{table}

\caption{  Predicted spectroscopic factors  in  the Nilsson scheme for the reactions $^{12}$Be$(d,p)$$^{13}$Be and $^{13}$B$(d,^3He)$$^{12}$Be.}
{\renewcommand{\arraystretch}{1.4}
\begin{tabular}{c|c|c|c|c}
\hline\hline
Initial & Final & Energy & \multirow{2}{*}{~$\ell$~}  & \multicolumn{1}{c}{Calculated $S_{i,f}$}\\
State & State & [MeV]  &            &  \\
\hline 
$^{12}$Be & $^{13}$Be&  &\\
 0$^{+}_1$ & $\frac{1}{2}^{+}$& 0.00&0 &0.52\\
                          & $\frac{5}{2}^{+}$& $\sim $1.8 &2 &0.47\\           
                          & $\frac{1}{2}^{-}$  & 0+x &1 &0.46\\         
\hline                     
$^{13}$B & $^{12}$Be& & & \\
 $\frac{3}{2}^{-}$ & 0$^{+}_1$ & 0.00&1&0.5\\
                            & 2$^{+}_1$ & 2.11&1&0.5\\
 & 0$^{+}_2$  & 2.24&1&0\\
                  \hline\hline
\end{tabular}
}

\label{Table2}
\end{table}

\section{Conclusion}
We have analyzed spectroscopic factors in $^{11}$Be and $^{12}$Be, obtained from $(d,p)$, $(-1n)$, and $(p,d)$ reactions, in the Nilsson strong coupling limit.
Using the formalism developed for studies of single-nucleon transfer reactions in deformed nuclei we derived, for the  cases considered,
simple formulae for spectroscopic factors in terms of the amplitudes of the deformed wavefunctions.  These amplitudes were empirically adjusted to reproduce the experimental data, 
including the magnetic moment of the $^{11}$Be ground state.  We have also used these wavefunctions to make some predictions for reactions such as $^{12}$B$(d,p)$$^{13}$Be and $^{13}$B$(d,^3He)$$^{12}$Be, that will likely be studied in the near future.

While more sophisticated microscopic approaches are available to describe the structure of neutron-rich $^{11}$Be and $^{12}$Be, their description in terms of a deformed mean-field
seems to capture the main physics ingredients~\cite{Ikuko}.  As shown in this work,  the approach also provides a satisfactory explanation of spectroscopic factors, in a simple and intuitive manner.

\section{Appendix}

We present here the formulae used to calculate the magnetic moment (see Ref.~\cite{BM}).
For a $K=1/2$ band the magnetic moment of the $I=1/2$ state is given by:

\begin{eqnarray*}
\mu=\frac{1}{2}g_R+\frac{g_K-g_R}{6}(1-2b)
\end{eqnarray*}
\noindent
where $g_R \approx Z/A$ and $g_K$  are the collective and single-particle gyromagnetic factors respectively, and $b$ is the magnetic decoupling parameter.

The gyromagnetic factor $g_K$ depends on the $C_{jl}$ amplitudes through the following relation:
\begin{eqnarray*}
g_K= g_s (C_{1/2,0}^2 + \frac{1}{5}(C_{5/2,2}^2 - C_{3/2,2}^2) - 2\sqrt{\frac{24}{25}} C_{5/2,2}C_{3/2,2})
\end{eqnarray*}
\noindent 
and the magnetic decoupling parameter $b$ is related to the decoupling parameter $a$:

\begin{eqnarray*}
b=   \frac{g_Ra - (g_s+g_K)/2}{(g_K-g_R)}
\end{eqnarray*}

\noindent 
with a:
\begin{eqnarray*}
a =   C_{1/2,0}^2 - 2 C_{3/2,2}^2 + 3C_{5/2,2}^2
\end{eqnarray*}

\noindent
Using the wavefunctions derived,  the calculated gyromagnetic factor $g_K$,  decoupling and magnetic-decoupling parameters for the ground state of $^{11}$Be are:
$g_K=-2.80$,  $a=1.92$ and $b=-1.27$ respectively.  We note that, associating the $5/2^{+}$ state at 1.78~MeV with the second member of the rotational band, its energy is given by:

\begin{eqnarray*}
E_{rot}= A( \frac{5}{2}( \frac{5}{2}+1) - a ( \frac{5}{2}+1))
\end{eqnarray*}

\noindent
With the rotational constant $A = 0.35$~MeV, determined from the  $2^{+}$ in $^{12}$Be, we estimate $a=1.85$, in excellent agreement with the value calculated from the magnetic moment.

\begin{acknowledgments}
 This material is based upon work supported by the U.S. Department of Energy, Office of Science, Office of Nuclear Physics under Contract No. DE-AC02-05CH11231 (LBNL). We thank Prof. R. Kanungo for thoughtful comments on the manuscript.\end{acknowledgments}

\end{document}